\UseRawInputEncoding
\documentclass[twocolumn]{aastex62}

\begin{document}

\title{Multi-band Emission up to PeV Energy from the Crab Nebula in a Spatially Dependent Lepto-hadronic Model }

\correspondingauthor{Li Zhang}
\email{lizhang@ynu.edu.cn}

\author{Qi-Yong Peng}
\affil{Department of Astronomy, Key Laboratory of Astroparticle Physics of Yunnan Province, Yunnan University, Kunming 650091, People's Republic of China}

\author{Bi-Wen Bao}
\affil{Department of Astronomy, Key Laboratory of Astroparticle Physics of Yunnan Province, Yunnan University, Kunming 650091, People's Republic of China}
\affil{Key Laboratory of Statistical Modeling and Data Analysis of Yunnan Province, Yunnan University, Kunming 650091, People's Republic of China}

\author{Fang-Wu Lu}
\affil{Department of Physics, Yuxi Normal University, Yuxi 653100, People's Republic of China}
\affil{Department of Astronomy, Key Laboratory of Astroparticle Physics of Yunnan Province, Yunnan University, Kunming 650091, People's Republic of China}

\author{Li Zhang}
\affiliation{Department of Astronomy, Key Laboratory of Astroparticle Physics of Yunnan Province, Yunnan University, Kunming 650091, People's Republic of China}

\begin{abstract}
Multi-band emission from radio to ultra-high energy gamma-rays in the Crab Nebula has been detected. To explain the observed results, non-thermal photon production \textbf{in} the Crab Nebula is carefully studied in a spatially dependent lepto-hadronic model. In our model, the dynamical evolution of the PWN is simulated in a spherically symmetric system. Both electrons and protons are accelerated at the termination shock. The relevant particle propagation equations as well as the photon evolving equation are simultaneously solved. For the Crab Nebula, our results reveal that the observed multi-band photon spectra can be well reproduced with reasonable model parameters. In particular, the photons with energy $\gtrsim 200$ TeV are mainly contributed by the hadronic component via proton-proton interaction. The contribution of the hadronic component depends on both proton spectral index $\alpha_{\rm p}$ and number density $n_{\rm H}$ of medium within the PWN. Besides, high energy neutrino fluxes are predicted with variable proton spectral indices. The predicted fluxes are not only far below the sensitivities of current neutrino observatories, but also beneath the atmospheric neutrino background with energy less than $\sim 40$ TeV. Moreover, the calculated radial profiles of surface brightness and spectral index are presented.
\end{abstract}


\section{Introduction}\label{sec:introduction}

The Crab Nebula, as the brightest representative of pulsar wind nebulae (PWNe), is a unique cosmic laboratory for probing the violent universe. Recently, LHAASO collaboration reported a discovery of $\sim $ PeV photon emission from the Crab Nebula, extending its energy spectrum to about 1.1 PeV with a power-law index $\sim 3$. The energy spectrum of the Crab Nebula now spans over 22 decades of energy, from radio to PeV gamma-ray band \citep{LHAASOCrab}. Such a striking discovery may spark renewed interest in this well-known source and has an important implication on extreme particle accelerators in the Galaxy.

Located at a distance of 2 kpc \citep{Manchester2005}, the Crab Nebula is a historical filled-center supernova remnant and has been carefully studied in great details. Present observations manifest several spectral features and a complex energy-dependent morphology regarding this specific source \cite[e.g.,][]{Hester2008,BB2014,YH19,Abdalla2020}. In its center, there exists a young pulsar (the Crab pulsar), rotating with a period of 33 ms and a period derivative of $4.23\times10^{-13} \rm s\ s^{-1}$. It is generally believed that the Crab Nebula is powered by its central pulsar through the formation and termination of ultra-relativistic electron-positron pulsar winds.

Theoretically, the non-thermal photons from PWNe are usually ascribed to pure leptonic models \cite[e.g.,][]{KC84,DH92,AA1996,ZCF2008,Torres2014,ZZF18,KAA20}. In these models, very and ultra-high energy (VHE and UHE) $\gamma$-rays originate from the inverse Compton scattering (ICS) of ultra-relativistic electrons/positrons. For the Crab Nebula, two distinct populations of electrons have been proposed to be responsible for the broad-band non-thermal emission \citep{AA1996}: "radio electrons" and "wind electrons". The former can produce MHz and GHz synchrotron photons, GeV $\gamma$-rays via ICS. The latter, also called as multi TeV electrons, is assumed to be injected at the pulsar wind termination shock and responsible for non-thermal photons at UV, X-ray, and TeV energy bands \citep{AA1996,ZCF2008,KAA20}.

Before the discovery of ultra-high-energy photons from the Crab Nebula by LHAASO\citep{LHAASOCrab}, the broad-band non-thermal emission of the Crab Nebula is well interpreted within a pure leptonic framework, while \textbf{hadronic models fail} to account for the overall broad-band $\gamma$-ray spectrum. However, in terms of PeV energies, the hadronic contribution due to p-p interaction may definitely not be marginal, as any $\gamma$-ray photons detected well beyond 1 PeV would require a non-leptonic origin \citep{LHAASOCrab}. The reasons for claiming such a statement are manifold: first of all, following the derivations in \citet{LHAASOCrab} (Eqs.1-4 therein), the maximum energy of electrons is limited by two conditions: (i) the electron gyro-radius cannot exceed the accelerator's linear size $l$. For the Crab Nebula, $l< 0.1 ~{\rm pc}$, (ii) the balance between the acceleration and energy loss rates. Thus, the maximum energy of electrons (photons) are constrained. Applying these conditions to the Crab Nebula, the maximum energy of photons produced via ICS cannot exceed several PeV (see Fig. S9 in \citet{LHAASOCrab}). Secondly, although acceleration at the TS could boost the energy of electrons to 1 PeV in principle, their subsequent escape from the acceleration site and further propagation inside the PWN (a magnetized region with nebular magnetic field $\sim 120 ~{\rm \mu G}$ as well as a radius of tens of parsecs) is a challenge.
Despite wide popularity of the pure leptonic model, the potential hadronic origin of UHE photons in the Crab Nebula cannot be ruled out \citep[e.g., ][]{AA1996,YZ09,ZCHC20,LHAASOCrab}, and such a hypothesis may further carry new implications for the origin of Galactic cosmic rays. Besides, much of the previous debates revolve around a pure leptonic origin, while there is insufficient research into this field to assess the potential contribution from a hadronic process.

In fact, as a young filled-center SNR, on one hand, this specific source could be a potentially powerful cosmic-ray accelerator, and the accelerated energetic protons may exert an influence on the resulting spectrum via hadronic interactions. On the other hand, the plausible pure leptonic models do not signify the absence of accelerated protons, but might only reveal that, under specific conditions (for instance, a low medium density $n_{\rm H}$ within the PWN), the hadronic component is less significant than its leptonic counterpart (but not negligible). Therefore, as a viable alternative to the pure leptonic origin, the lepto-hadronic model deserves close scrutiny in a detailed way. As for specific scenarios in explaining SEDs of PWNe, the simplest models are spatially-independent \cite[e.g.,][]{AA1996,ZCF2008,Gelfand2009,Torres2014,KAA20,ZCHC20}. Recently, a spatially-dependent model has been proposed, with applications to several PWNe \cite[e.g.,][]{Lu2017,Lu2019,Lu2020}.

In this paper, a spatially-dependent lepto-hadronic model is proposed and applied to explain the SED of the Crab Nebula, especially its PeV emission with a potential hadronic origin. This paper proceeds as follows. A detailed description of our spatially-dependent model is in Section 2. The calculated results and a direct comparison with the observed data are presented in Section 3. Finally, some discussion and conclusions are given in Section 4.

\section{Model Descriptions}\label{sec:model}

In our lepto-hadronic model, for simplicity, three assumptions are made: (i) in the PWN, hadrons and leptons correspond to protons and electrons respectively, and both species are energized by the central pulsar.  \citep[e.g.,][]{Horns2006,ZY2009}, (ii) particle injection occurs at the termination shock (TS) \citep[e.g., ][]{KC84}, and (iii) a spherically symmetric system is utilized while the model proposed by \citet{Bucciantini2011} is adopted to simulate the dynamical evolution of the PWN (see also \citet{Lu2017}).

\subsection{Particle Energy Input in a PWN} \label{subsec:energysource}

The central pulsar provides energy for a PWN through its spin-down power. The spin-down power is given by:
\begin{equation}
L(t)=L_0\left(1+\frac{t}{\tau_0}\right)^{-\frac{n+1}{n-1}}\;,
\label{Lsd}
\end{equation}
where $L_0$ is the initial spin-down power, $n$ is the braking index ($n = 3$ for a pure dipole radiative pulsar). $\tau_0$ is the initial spin-down timescale, which can be expressed as
\begin{equation}
\tau_0=\frac{2}{n-1}\tau_{\rm c} - T_{\rm age}\;,
\end{equation}
where $\tau_{\rm c}=P/[(n-1)\dot{P}]$ is the characteristic age while $T_{\rm age}$ is the pulsar age ($P$ is the pulsar period and $\dot{P}$ is the period derivative).

\textbf{According to PWNe MHD models}, the majority of the spin-down power is released in a relativistic particle wind \citep[e.g., ][]{KC84}. These particles are injected and accelerated at the termination shock (TS), with their injection rates being
\begin{equation}\label{Qinjp}
Q_{\rm inj}^{\rm p}(\gamma_{\rm p},t)=Q^{\rm p}_0(t){\gamma_{\rm p}}^{-\alpha_{\rm p}} e^{-(\gamma_{\rm p}/\gamma_{\rm p,c})^2}~~{\gamma_{\rm p,min}}\le \gamma_{\rm p}\le \gamma_{\rm p,max}
\end{equation}
for protons and
\begin{equation}
Q_{\rm inj}^{\rm e}(\gamma_{\rm e}, t)=Q^{\rm e}_0(t)
\left\{\begin{array}{ll}
\left(\frac{\gamma_{\rm e}}{\gamma_{\rm b}}\right)^{-\alpha_1}~~\gamma_{\rm e,min}\le \gamma_{\rm e}<\gamma_{\rm b} \\
\left(\frac{\gamma_{\rm e}}{\gamma_{\rm b}}\right)^{-\alpha_2}~~{\gamma_{\rm b}}\le \gamma_{\rm e}\le \gamma_{\rm e,max}
  \end{array}
  \right.
  \label{Qinje}
\end{equation}
for electrons, respectively. In above equations, $Q^{\rm p}_0$ and $\gamma_{\rm p}$ ($Q^{\rm e}_0(t)$ and $\gamma_{\rm e}$) are the normalization coefficient as well as the Lorentz factor of protons (electrons), $\gamma_{\rm p,c}$ is the proton's cutoff Lorentz factor, $\gamma_{\rm b}$ is the electron's break Lorentz factor, \textbf{and} $\gamma_{\rm p,min}$ and $\gamma_{\rm p,min}$ ($\gamma_{\rm e,min}$ and $\gamma_{\rm e,min}$) are the minimum and maximum Lorentz factors of protons (electrons).

In above equations, the maximum energy of protons (electrons) is a key physical quantity. To gain energy, particles must be confined inside the region with radius $\lesssim$ $R_{\rm ts}= (1/B_1)\sqrt{(L(t)/c)\sigma/(1+\sigma)}$ \citep{KC84}, where $B_1$ is upstream magnetic field, $L(t)$ is the spin-down power given by Eq. (\ref{Lsd}), $c$ is the light speed, and $\sigma$ is the magnetization parameter. Therefore, the particle's Larmor radius, $r_{\rm L}=pc/(ZeB)$ should not be larger than $R_{\rm ts}$ ($B$ is downstream magnetic field). Using the condition of $r_{\rm L}=\varepsilon R_{\rm ts}$ with $0<\varepsilon\le 1$, the maximum energy of protons (or electrons) can be approximated as
\begin{equation}
\gamma_{i,{\rm max}}\approx \frac{4\varepsilon e}{m_ic^2}\sqrt{\frac{L(t)}{c}\frac{\sigma}{1+\sigma}}= \frac{4\varepsilon e}{m_ic^2}\sqrt{\eta_{\rm B}\frac{L(t)}{c}}\;,
\label{Gammax}
\end{equation}
where $i= \rm e, \rm p$ denotes electrons and protons, respectively. $e$ is the electron charge, and $\eta_{\rm B}=\sigma/(1+\sigma)$ describes the fraction of the magnetic energy converted from the spin-down luminosity. For the Crab Nebula, the maximum energy of the protons is given by
\begin{equation}
\frac{E_{\rm p, max}}{\rm PeV} \approx 18\left(\frac{\varepsilon}{0.7}\right)\left(\frac{\eta_{\rm B}}{0.03}\right)^{\frac{1}{2}}\left(\frac{L(t)}{5\times 10^{38}~{\rm erg/s}}\right)^{\frac{1}{2}}\;.
\label{Gammax1}
\end{equation}
Obviously, the value of $\eta_{\rm B}$ (or $\sigma$) is important for estimating $E_{\rm p, max}$. In fact, the simple one-dimensional model has $\sigma = 3\times 10^{-3}$ \citep{KC84}, which gives $E_{\rm p, max}\sim 6$ PeV; while axisymmetric two-dimensional simulations reveal that $\sigma\gtrsim 0.03$ \citep{DVAB06}, resulting in $E_{\rm p, max}\gtrsim 18$ PeV. More recently, three-dimensional simulations show that the value of $\sigma$ can be larger than 1 \citep[e.g.,][]{PKK14,Amato19}, which corresponds to $\eta_{\rm B}=0.5$ and $E_{\rm p, max}\sim 76$ PeV.

Using Eqs. (\ref{Qinjp}) and (\ref{Qinje}), the energy per unit time $W_{\rm p}$ (for protons) and $W_{\rm e}$ (for electrons) are
$W_{\rm p}=\int\gamma_{\rm p} m_{\rm p}c^2Q_{\rm inj}^{\rm p}(\gamma_{\rm p},t)d\gamma_{\rm p}$ and $W_{\rm e}=\int\gamma_{\rm e} m_{\rm e}c^2Q_{\rm inj}^{\rm e}(\gamma_{\rm e},t)d\gamma_{\rm e}$, respectively. Since the particle energy is converted from the spin-down power $L(t)$, then $W_{\rm p}=\eta_{\rm p} L(t)$ and $W_{\rm e}=\eta_{\rm e}L(t)$.
$\eta_{\rm p}$ ($\eta_{\rm e}$) is the energy fraction occupied by protons(electrons), which is a free parameter in our calculations. Thus, the normalization coefficients of protons and electrons are given by
\begin{equation}\label{Q0p}
Q_0^{\rm p}(t) =\frac{\eta_{\rm p}L(t)}{m_{\rm p}c^2}\left[\int_{\gamma_{\rm p,min}}^{\gamma_{\rm p,max}}{\gamma_{\rm p}}^{-\alpha_{\rm p}+1}e^{-\frac{\gamma_{\rm p}}{\gamma_{\rm p,c}}}d\gamma_{\rm p}\right]^{-1}\;,
\end{equation}
and
\begin{eqnarray}\label{Q0e}
&&Q_0^{\rm e}(t) =\frac{\eta_{\rm e}L(t)}{\gamma_{\rm b}m_{\rm e}c^2}\\\nonumber
&&\left[\int_{\gamma_{\rm e,min}}^{\gamma_{\rm b}}\left(\frac{\gamma_{\rm e}}{\gamma_{\rm b}}\right)^{-\alpha_1+1}d\gamma_{\rm e} + \int^{\gamma_{\rm e,max}}_{\gamma_{\rm b}}\left(\frac{\gamma_{\rm e}}{\gamma_{\rm b}}\right)^{-\alpha_2+1}d\gamma_{\rm e}\right]^{-1}\;.
\end{eqnarray}
Note that $\eta_{\rm p}+ \eta_{\rm e} +\eta_{\rm B}=1$.

\subsection{Particle Transport Equations} \label{subsec:TransportEquations}

In a spherically symmetric system,  the transport of particles with  number density $n_{i }\equiv n_{i}(r,\gamma,t)$ within the nebula can be written as  \citep[e.g.,][]{Vorster2013a,Lu2017}.
\begin{eqnarray}
\nonumber
\frac{\partial{n_{i}}}{\partial{t}}
&=&D_i\frac{\partial^2n_{i}}{\partial{r}^2}+
\left[\frac{1}{r^2}\frac{\partial}{\partial{r}}(r^2D_i)-V\right]\frac{\partial{n_{i}}}{\partial{r}}\\
&&-\frac{1}{r^2}\frac{\partial}{\partial{r}}\left[r^2V\right]n_{i}+\frac{\partial}{\partial{\gamma_i}}\left[\dot{\gamma_i}n_{i}\right]+ Q_{i}\;,
\label{ni}
\end{eqnarray}
where $i={\rm e,~p}$ denotes electrons and protons respectively, $V=V(r)$ is the bulk velocity, $D_i\equiv D_i(r,\gamma_i,t)$ is the diffusion coefficient of particles, $\dot{\gamma}_i$ is the summation of particle energy losses, and $ Q_{i}$ is the source term. These physical quantities are described below.

In general, the evolution of a PWN could be roughly divided into three stages, that is, free expansion phase, reverberation phase, and Sedov-Taylor phase \citep[e.g.][]{Gaensler2006,Gelfand2009,Bucciantini2011,Kolb2017}. In the free expansion phase, the temporal evolution of the radius of the PWN is described by
\begin{equation}
R_{\rm pwn}(t)=R_0(t)\left(\frac{1}{1+\frac{t}{\tau_0}}\right)^{\frac{6}{5}}\frac{1}{1-s}\sum_{i=0}^{\infty}c_is^i\;,
\label{Free}
\end{equation}
where the terms $R_0(t)=(E_{\rm SN}^3L_0^2/M_{\rm ej}^5)^{\frac{1}{10}}t^{\frac{6}{5}}$, $s=(t/\tau_0)/(1+t/\tau_0)$, and the coefficient $c_i$ can be obtained from \citet{Bucciantini2004}. Initial parameters, including the ejecta mass $M_{\rm ej}$ and the explosion energy $E_{\rm SN}$, are listed in Table \ref{tab:1}. Once the nebula has reached the SNR's reverse shock, it collides with the shocked ejecta and then the reverberation phase triggers \citep[][]{vanderSwaluw2001}. The radius of the nebula at this stage is calculated by
\begin{equation}
M_{\rm sw}\frac{d^2{R}_{\rm pwn}(t)}{dt^2}=4\pi R^2_{\rm pwn}(t)[P_{\rm in}(t)-P_{\rm out}(t)]\;,
\label{rever}
\end{equation}
where $P_{\rm in}$ and $P_{\rm out}$ are the pressures of the PWN shell on the inner and outer sides respectively, while $M_{\rm sw}$ is the cumulative medium mass swept up by the nebula.
When the pressure inside the PWN meets the Sedov solution and the swept-up material begins to dominate the dynamics of the SNR \citep[e.g.][]{Bucciantini2011}, the Sedov-Taylor evolutionary phase starts. Then, the radius of the nebula is determined by
\begin{equation}
R^4_{\rm pwn}(t_{\rm Sedov})P_{\rm in}(t_{\rm Sedov})=R^4_{\rm pwn}(t)P(t)\;,
\label{Sedov}
\end{equation}
where $t_{\rm Sedov}$ is the time at which the Sedov-Taylor phase begins, and $P(t)=n_{\rm ism}v^2_{\rm fs}/(\gamma_{\rm snr}+1)$ is the pressure at the SNR forward shock. Here, $v_{\rm fs}$ is the speed of the forward shock, $\gamma_{\rm snr}$ is the adiabatic index, while $n_{\rm ism}$ is the density of the surrounding interstellar medium (ISM). Finally, the dynamical radius of the TS can be estimated by
\begin{equation}
R_{\rm ts}(t)=\sqrt{\frac{L(t)}{4{\pi}cP_{\rm ts}(t)}}\;,
\label{RTS}
\end{equation}
where $P_{\rm ts}(t)$ is the pressure of the gas at the TS. Regarding the evolution of a non-radiative SNR, we use the equations collected from \citet{Gelfand2009} to describe the temporal evolutions of the forward shock($R_{\rm snr}$) as well as the reverse shock ($R_{\rm rs}$).

\begin{figure}
\figurenum{1}
\includegraphics[width=0.50\textwidth,angle=0]{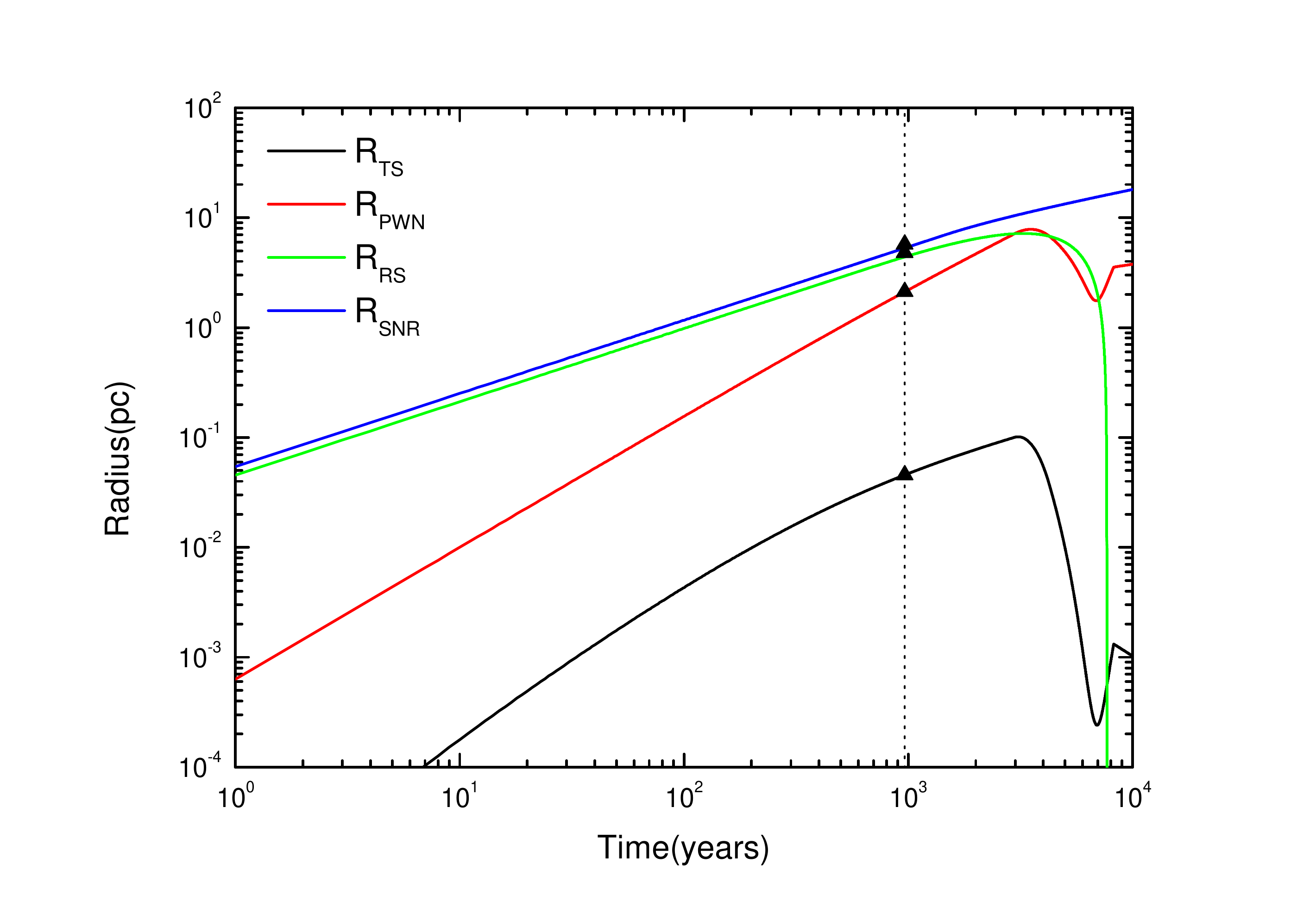}
\caption{Temporal evolutions of the forward shock ($R_{\rm snr}$), the reverse shock ($R_{\rm rs}$), the PWN ($R_{\rm pwn}$), and the TS ($R_{\rm ts}$). The vertical dashed line represents the current age of the Crab Nebula ($966~{\rm yr}$).}
\label{fig:1}
\end{figure}

With the input parameters listed in Table \ref{tab:1}, \textbf{the temporal evolutions of the forward shock($R_{\rm snr}$), the reverse shock ($R_{\rm rs}$), the PWN ($R_{\rm pwn}$), and the TS ($R_{\rm ts}$), are calculated and displayed in Fig. \ref{fig:1}.} As shown in the figure, $R_{\rm pwn}\approx 2.1$ pc; while $R_{\rm ts}\approx 0.05$ pc, less than the upper limit $R_{\rm ts}\lesssim 0.1$ pc deduced from X-ray observations \citep{W2000}.

For the bulk velocity of particles, its radial profile follows $V\propto 1/r^2$ at the nebula inner region and approaches a constant as the nebula radius increases to a hundred times of the TS radius \citep{KC84}. Without loss of generality, $V\propto r^{-\beta}$ is assumed ($\beta=2$ for the Crab Nebula). If the bulk velocity at the outer edge of the nebula is a constant and could be characterized by $R_{\rm pwn}(T_{\rm age})/T_{\rm age}$, where $R_{\rm pwn}(T_{\rm age})$ is the nebula radius while $T_{\rm age}$ is the age of the PWN, then the bulk velocity can be expressed as
\begin{equation}\label{Vr}
V(r)=\frac{R_{\rm pwn}(T_{\rm age})}{T_{\rm age}}\left(\frac{r}{R_{\rm pwn}(t)}\right)^{-\beta}\;,
\end{equation}
where $R_{\rm pwn}(t)$ is the nebula radius at time $t$.

The radial profile of the magnetic field inside the nebula can be given by
\begin{equation}
B(r,t)=B_0(t)\left(\frac{r}{R_{\rm ts}(t)}\right)^{\beta-1}\;,
\end{equation}
where $B_0(t)$ is the magnetic field at the TS, which can be obtained by solving the following equation \citep[e.g.,][]{Pacini1973}:
\begin{equation}
\frac{dW_{\rm B}(t)}{dt}=\eta_{\rm B}L(t)-\frac{W_{\rm B}(t)}{R_{\rm pwn}(t)}\frac{dR_{\rm pwn}(t)}{dt}\;,
\label{Mag}
\end{equation}
where $W_{\rm B}(t)=\int{B^2(r,t)r^2dr/2}$.

The particle diffusion coefficient, which relates to magnetic field as $\propto1/B(r,t)$ \citep[][]{CaballeroLopez2004} and particle energy as $\propto\gamma_i^{\delta}$ with $\delta=1/3$ \citep[][]{Aguilar2016}, can be expressed as
\begin{equation}
D(r,\gamma_i,t)=D_0(t)\left(\frac{r}{R_{\rm ts}(t)}\right)^{1-\beta}\left(\frac{E_i}{1{\rm TeV}}\right)^{1/3}\;,
\label{diff}
\end{equation}
where $E_i=\gamma_im_ic^2$ is the particle energy and $m_i$ represents the mass of particles.

Since the energy loss and source term for protons are different from those for electrons, they are described respectively as follows.

\subsubsection{Proton energy loss rate and source term}

For protons, the total energy loss includes adiabatic loss $\dot{\gamma}_{\rm ad}$ and the radiative loss $\dot{\gamma}_{\rm pp}$ of p-p interaction, i.e.,
\begin{eqnarray}
\dot{\gamma}_{\rm p}=\dot{\gamma}_{\rm ad} + \dot{\gamma}_{\rm pp}\;.
\label{gammadot}
\end{eqnarray}
The adiabatic loss can be expressed as
\begin{eqnarray}\label{addot}
\dot{\gamma}_{\rm ad}(r,\gamma_{\rm p},t)=\frac{1}{3}\nabla{\cdot{\textbf{\textit{V}}}}\gamma_{\rm p}= \frac{1}{3}\left[\frac{2V}{r}+\frac{\partial V}{\partial r}\right]\gamma_{\rm p}\;,
\end{eqnarray}
where $V$ is the convection velocity and $\gamma_{\rm p}=E_{\rm p}/m_{\rm p}c^2$. The energy loss in p-p interaction is given by
\begin{eqnarray}
\nonumber
\dot{\gamma}_{\rm pp}(r,E_{\rm \gamma},t)&=& cn_{\rm H}\int_{E_{\rm \gamma}}^{\infty}\sigma_{\rm inel}(E_{\rm p})\times\\
&&n_{\rm p}(r,\gamma_{\rm p},t)F_{\gamma}(E_{\rm \gamma}/E_{\rm p},E_{\rm p})\frac{dE_{\rm p}}{E_{\rm p}}\;,
\end{eqnarray}
where $n_{\rm H}$ is the hydrogen number density in PWN medium, $\sigma_{\rm inel}(E_{\rm p})$ is the inelastic cross section, $E_\gamma$ is the photon energy and $F_{\gamma}(E_{\rm \gamma}/E_{\rm p},E_{\rm p})$ is the gamma-ray production spectrum \citep[for details, see][]{Kelner2006}.

The source term of protons is $Q_p(\gamma_{\rm p}, t)= Q_{\rm inj}^{\rm p}(\gamma_{\rm p},t)$, which is given by Eq. (\ref{Qinjp}).

\subsubsection{Electron energy loss rate and source term}

The total energy loss of electrons consists of adiabatic loss $\dot{\gamma}_{\rm ad}$, synchrotron radiation $\dot{\gamma}_{\rm syn}$, and IC scattering $\dot{\gamma}_{\rm ics}$, which is given by
\begin{eqnarray}
\dot{\gamma}_{\rm e}=\dot{\gamma}_{\rm ad}+\dot{\gamma}_{\rm syn}+\dot{\gamma}_{\rm ics}\;.
\label{gammadot1}
\end{eqnarray}
In above expression, the adiabatic loss is described by Eq. (\ref{addot}), where $\gamma_{\rm p}$ is replaced with $\gamma_{\rm e}=E_{\rm e}/m_{\rm e}c^2$; the synchrotron radiation loss is given by \citep{Rybicki1979}
\begin{eqnarray}
\dot{\gamma}_{\rm syn}(r,\gamma_{\rm e},t)=\frac{4}{3}\frac{\sigma_{\rm T}}{m_{\rm e}c}\gamma^2_{\rm e}U_{\rm B}(r,t)\;,
\end{eqnarray}
where $\sigma_{\rm T}$ is the Thomson cross section and $U_{\rm B}(r,t)=B^2(r,t)/8\pi$ is the magnetic field energy density; the IC loss $\dot{\gamma}_{\rm ics}$ is given by \citep{Blumenthal1970}
\begin{eqnarray}
\nonumber
\dot{\gamma}_{\rm ics}(r,\gamma_{\rm e},t)&=&\frac{3}{4}\sigma_{\rm T} c \frac{1}{\gamma^2_{\rm e}}\int_0^{\infty}\epsilon_fd\epsilon_f\int_0^{\infty}\frac{n_{\rm sph, j}(r,\epsilon_i,t)}{\epsilon_i}\\
&&\times f(q,\Gamma) H(1-q)(q-\frac{1}{4\gamma^2})d\epsilon_i\;,
\end{eqnarray}
where $n_{\rm sph, j}(r, \epsilon_i,t)$ is the number density of the soft photon fields with initial photon energy $E_{\gamma,i}=m_{\rm e} c^2\epsilon_i$, $\epsilon_i$ and $\epsilon_f$ are the initial and final energies of the scattered photons, respectively. $H$ is the Heaviside step function, with $\Gamma=4\gamma_{\rm e} \epsilon_i$, $q=\epsilon_f/(\Gamma(\gamma_{\rm e}-\epsilon_f))$, and $f(q,\Gamma)=2q\ln q+(1+2q)(1-q)+0.5(1-q)(\Gamma q)^2/(1+\Gamma q)$. Note that the soft photon fields include thermal ($n_j(\epsilon_i)$) and non-thermal photon ($n_{\rm syn}(r, \epsilon_i, t)$) components, i.e., $n_{\rm sph, j}=n_j +n_{\rm syn}$. Specifically, there are three thermal photon fields: cosmic microwave background (CMB), infrared (IR), and NIR/optical emission from nearby stars. Each field contains an energy density $U_j$ and a temperature $T_j$, with the subscript $j$ representing one of the three components. The thermal photon number density here \textbf{is} given by
\begin{eqnarray}
\nonumber
n_{j}(h\nu_i)=\frac{15U_j}{(\pi kT_j)^4}\frac{(m_ec^2\epsilon_i)^2}{\exp(m_ec^2\epsilon_i/kT_j)-1}\;.
\end{eqnarray}
On the other hand, the non-thermal soft photons are the synchrotron photons and their number density is
\begin{eqnarray}
\nonumber
n_{\rm syn}(r, \epsilon_i, t)=\frac{Q_{\rm syn}(\epsilon_i)}{4\pi c R^2_{\rm ts}(t)}U(x)\;,
\end{eqnarray}
with
\begin{eqnarray}
\nonumber
U(x)=\frac{3}{2}\int_0^1\frac{y}{x}\ln\frac{x+y}{\mid x-y \mid}dy\;,
\end{eqnarray}
where $x\equiv r/R_{\rm ts}(t)$.

As to the source term of electrons, it includes two terms, i.e.,
\begin{equation}\label{Qeinj}
Q_{\rm e, \rm inj} = Q_{\rm inj}^{\rm e} + Q_{\rm pp}^{\rm e}\;,
\end{equation}
where the first term on the right side is derived by Eq. (\ref{Qinje}), while the second term is given by
\begin{eqnarray}
\nonumber
Q_{\rm pp}^{\rm e}(r, E_{\rm e}, t) &= cn_{\rm H}\int_{E_{\gamma}}^{\infty}\sigma_{\rm inel}(E_{\rm p})n_{\rm p}(r,E_{\rm p},t)\\
&\times F_{\rm e}(E_{\rm e}/E_{\rm p}, E_{\rm p})\frac{dE_{\rm p}}{E_{\rm p}}\;,
\end{eqnarray}
with $F_{\rm e}(E_{\rm e}/E_{\rm p},E_{\rm p})$ being the electron spectrum via the process $\pi \to \mu\nu_\mu$ \citep[for details, see][]{Kelner2006}.

\subsection{Photon Evolving Equation}

During the particle propagation, photons are produced through various interaction processes. \textbf{The photon number density $n_{\gamma}$ as a function of position $r$, energy ($\epsilon \equiv h\nu/(m_{\rm e}c^2)$), and time evolves following}
\begin{eqnarray}
\frac{\partial{n_{\gamma}}}{\partial{t}}+\frac{n_{\gamma}}{\tau_{\rm esc}}=
Q_{\gamma}^{\rm syn}+Q_{\gamma}^{\rm ics}+Q_{\gamma}^{\rm pp}\;,
\label{nx}
\end{eqnarray}
where $\tau_{\rm esc}=R_{\rm pwn}(t)/c$ is the photon escape timescale, $Q_{\gamma}^{\rm syn}$ and $Q_{\gamma}^{\rm ics}$ are the emissivity of the synchrotron radiation and IC scattering of the electrons, while $Q_{\gamma}^{\rm pp}$ is the emissivity of photons in p-p interaction.

The synchrotron emissivity is given by \citep{Blumenthal1970}
\begin{eqnarray}
Q_{\gamma}^{\rm syn}(r, \epsilon, t)=\int_0^{\infty}n_{\rm e}(r,\gamma,t)P_{\rm syn}(r, \epsilon, \gamma, t)d{\gamma}\;,
\end{eqnarray}
where $P_{\rm syn}(r, \epsilon, \gamma, t)$ is the single particle synchrotron emissivity:
\begin{eqnarray}
P_{\rm syn}(r, \epsilon, \gamma, t)=\frac{\sqrt{3}e^3B(r,t)}{h\epsilon m_{{\rm e}}c^2}F\left(\frac{\epsilon}{\epsilon_{\rm c}}\right)\;,
\end{eqnarray}
where $\epsilon_{\rm c}=3heB(r,t)\gamma^2/(4{\pi}m_{\rm e}^2c^3)$, $F({\rm y})={\rm y}\int_{\rm y}^{\infty}K_{5/3}({\rm z})d{\rm z}$ with ${\rm y}=\epsilon/\epsilon_{\rm c}$, and $K_{5/3}$ is a modified Bessel function of order $5/3$.

The emissivity of IC scattering in the Klein-Nishina regime is described by \citep{Blumenthal1970}
\begin{eqnarray}
Q_{\gamma}^{\rm ics}(r, \epsilon, t)=\int_0^{\infty}n_{\rm e}(r,\gamma,t)P_{\rm IC}(r, \epsilon, \gamma, t)d{\gamma}\;,
\end{eqnarray}
where $P_{\rm IC}(r, \epsilon, \gamma, t)$ is the scattered photon emissivity per electron:
\begin{eqnarray}
\nonumber
P_{\rm IC}(r,\epsilon, \gamma, t)&=&\frac{3}{4}\frac{\sigma_{\rm T}c}{\gamma^2}\int_0^{\infty}\frac{n_{\rm sph, j}(r,\epsilon_i,t)}{\epsilon_i}\\
&&\times f(q,\Gamma)d{\epsilon_i}\;.
\end{eqnarray}

The emissivity of photons in p-p interaction is expressed as
\begin{eqnarray}\nonumber
    Q_{\gamma}^{\rm pp}(r, E_{\rm \gamma}, t) &=& cn_{\rm H}\int_{E_{\gamma}}^{\infty}\sigma_{\rm inel}(E_{\rm p})\\
   &&\times n_{\rm p}(r,E_{\rm p},t)F_{\gamma}(E_{\rm \gamma}/E_{\rm p}, E_{\rm p})\frac{dE_{\rm p}}{E_{\rm p}}\;,
    \label{Qppgama}
\end{eqnarray}
where $\sigma_{\rm inel}(E_{\rm p})$ is the inelastic cross section, and $F_{\gamma}(E_{\rm \gamma}/E_{\rm p},E_{\rm p}) $ is $\gamma$-ray production spectrum \citep[for details, see][]{Kelner2006}.

\textbf{The spatially averaged spectral energy distribution in the PWN is given by
\begin{eqnarray}
E^2\frac{dN}{d E}&=&\frac{\epsilon^2 m_{{\rm e}}c^2}{4\pi d^2 }\int_{R_{\rm ts}}^{r} 4\pi r_{1}^{2}n_{\gamma}(r_1,\epsilon,t)\frac{dr_1}{\tau'_{\rm esc}}\;.
\end{eqnarray}
where $\tau'_{\rm esc}=(r-R_{\rm ts})/c$, $d$ is the distance from Earth to the system and $R_{\rm ts}$ is the radius of the termination shock.}

The surface brightness at different positions are calculated according to \citet{Holler2012}.


\begin{table}
\begin{centering}
\caption{Values of parameters for the Crab Nebula.\label{tab:1}}
\begin{tabular}{ccc}
\hline
\hline
Input parameters & Symbol & Value\\
\hline
Ejected mass($\rm{M_\odot}$)    & $M_{\rm ej}$  &   4.5\\
SN explosion energy ($10^{51}~\rm{erg}$)   & $E_{\rm SN}$ &   1.0 \\
Period ($\rm{ms}$)   &   $P$     &  33.04\\
Period derivative ($\rm{s\cdot s^{-1}}$)   &   $\dot{P}$    &   $4.23\times 10^{-13}$ \\
Initial spin-down power ($\rm{erg~s^{-1}}$)  & $L_0$  &   $3.0\times10^{39}$\\
Initial spin-down timescale (yr)     &   $\tau_0$   &   705.0       \\
Braking Index & $n$ & 2.509\\
Age (yr) & $T_{\rm age}$ & 966\\
Distance (kpc) &  $d$  &   2.0\\
Shock radius fraction  &  $\varepsilon$ &  0.7 \\
\hline
Fitted parameters & & \\
\hline
Magnetic fraction    & $\eta_{\rm B}$        & 0.06  \\
Electron fraction    & $\eta_{\rm e}$        & 0.70  \\
Proton fraction    & $\eta_{\rm p}$             & 0.24  \\
Proton cutoff energy (PeV)     &$E_{\rm p,c}$               &10\\
Low energy power-law index & $\alpha_1$ & 1.5\\
High energy power-law index & $\alpha_2$ & 2.5\\
Break Lorentz factor   & $\gamma_{\rm b}$ & $5.0\times10^5$  \\
Initial diffusion coefficient ($\rm cm^2~s^{-1}$) & $D_0$ & $1.0\times10^{22}$\\
\hline
\end{tabular}
\end{centering}
\end{table}

\subsection{The Neutrino Flux} \label{subsec:Nflux}

For the muonic neutrinos produced in the nebula, predicted spectrum of muonic neutrinos detected on the Earth is
\begin{equation}\label{neutrinoplux}
E^2_{\nu_\mu}\frac{dN}{dE_{\nu_\mu}} = \frac{E^2_{\nu_\mu}}{4\pi d^2}\int^r_{R_{\rm ts}} 4\pi r^2_1 Q_{\nu_\mu}(r_1, E_{\nu_\mu})dr_1
\end{equation}
where $Q_{\nu_\mu}(r_1, E_{\nu_\mu})$ is the emissivity of the muonic neutrinos, which is given by
\begin{eqnarray}
&&Q_{\nu_\mu}(r, E_{\nu_\mu}) = cn_{\rm H}\int^\infty_{E_{\nu_\mu}}\sigma_{\rm inel}(E_{\rm p})n_{\rm p}(r, E_{\rm p})\\\nonumber
&&\times [F_{\nu^1_\mu}(E_{\nu_\mu}/E_{\rm p}, E_{\rm p}) + F_{\nu^2_\mu}(E_{\nu_\mu}/E_{\rm p}, E_{\rm p})]\frac{dE_{\rm p}}{E_{\rm p}}\;,
\end{eqnarray}
where $F_{\nu^i_\mu}(E_{\nu_\mu}/E_{\rm p}, E_{\rm p})$ ($i=1,2$) is the neutrino production spectrum \citep[for details, see][]{Kelner2006}.

\subsection{Calculation Setups} \label{subsec:Setup}

After detailing our lepto-hadronic model, the boundary conditions are required to solve Eqs. (\ref{ni}) and (\ref{nx}).

Since the number of particles flowing into the nebula shall be equal to the number of particles injected at the TS, the inner boundary condition located at the TS for Eq. (\ref{ni}) should satisfy the following form:
\begin{equation}
V_0n_i - D_i(R_{\rm ts},\gamma_i,t)\frac{\partial{n_i}}{\partial r} = \frac{Q_{i,\rm inj}}{4\pi R^2_{\rm ts}(t)}\;,
\end{equation}
where $V_0$ is the velocity at the TS. On the other hand, to simulate the particles escaping from the PWN, a free escape condition is imposed at the outer boundary \citep{Vorster2013a}: $n_i (R_{\rm pwn}, \gamma_{\rm i}, t) = 0$ with $i={\rm e,~p}$. Under the above boundary conditions, transport equations for protons and electrons (Eq. (\ref{ni})) are solved numerically by using an
Alternating Direction Implicit method \citep{Douglas1962}. For Eq. (\ref{nx}), following the imposed free escape condition, it is numerically solved by using the Crank-Nicolson method \citep{Crank1996}.

\section{Application to the Crab Nebula} \label{sec:results}

In this section, the lepto-hadronic model elucidated above is applied to the Crab Nebula. As a famous PWN, the Crab Nebula retains a typical age of 966 years and a measured braking index $n=2.509$ \citep{Lyne1993}. The chosen explosion energy and ejecta mass are $E_{\rm SN}=1.0\times 10^{51}$ ergs and $M_{\rm ej}=4.5M_\odot$, respectively, while the density of ISM $n_{\rm ism}$ is assumed to be $0.2$ cm$^{-3}$.

In our calculations, \textbf{the background photon fields involved in the IC scattering process are as follows. For CMB, a temperature $T_{\rm CMB}=2.73~{\rm K}$ and an energy density $U_{\rm CMB}=0.26~{\rm eV~cm^{-3}}$ are utilized. For NIR, $T_{\rm NIR}=3000~{\rm K}$ and $U_{\rm NIR}=0.1~{\rm eV~cm^{-3}}$ are adopted parameters, while for FIR,
$T_{\rm FIR}=25~{\rm K}$ and $U_{\rm FIR}=0.9~{\rm eV~cm^{-3}}$ are used.} On the other hand, the proton spectral index $\alpha_{\rm p}$ and the number density $n_{\rm H}$ of medium within the PWN (i.e. target gas density for hadronic process) constitute a combination.

According to our results, the current magnetic field profile in the Crab Nebula seems to manifest a radially decreasing trend, that is, the magnetic field decreases from $217.0~{\rm \mu G}$ at the TS to $22.6~{\rm \mu G}$ at the outer edge of the nebula. Besides, the diffusion coefficient of particles at $1~{\rm TeV}$ varies from $5.6\times10^{27}~{\rm cm^2~s^{-1}}$ to $5.4\times10^{28}~{\rm cm^2~s^{-1}}$, with the increasing radial distance from the TS to the outer boundary. The current spatially averaged magnetic field and diffusion coefficient are estimated to be $62.6~{\rm \mu G}$ and $2.9\times 10^{28}~{\rm cm^2~s^{-1}}$, respectively.

\begin{figure}
\figurenum{2}
\includegraphics[width=0.50\textwidth,angle=0]{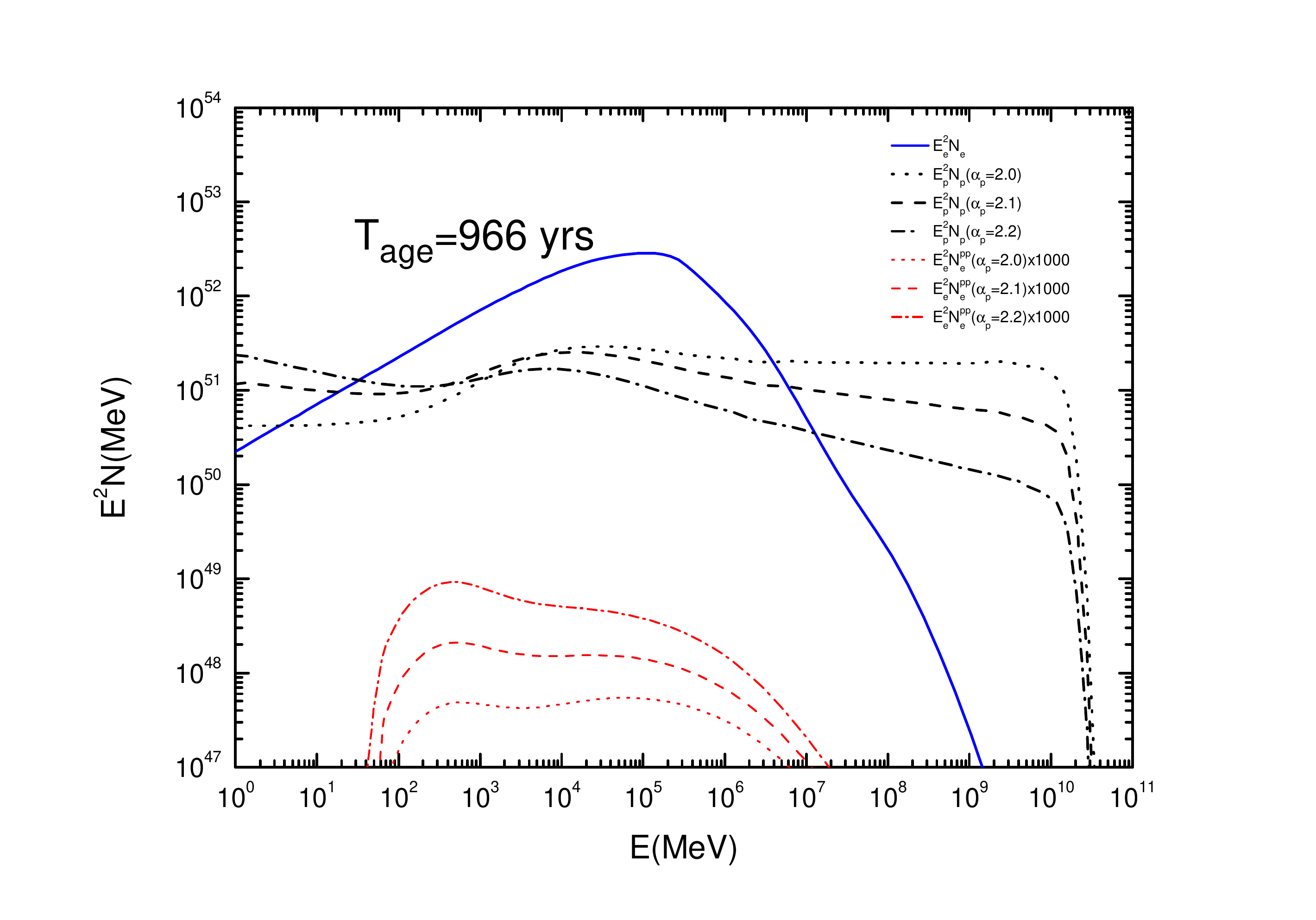}
\caption{Electron and proton spectra with $\alpha_{\rm p}$: $2.1$, $2.2$, and $2.3$ for the Crab Nebula. The parameters adopted here are listed in Table \ref{tab:1}.
  The blue solid line represents the spectrum of primary electrons, while the dashed (dotted and dot-dashed) red lines are for secondary electrons. The dashed (dotted and dot-dashed) black lines stand for proton spectra with different spectral indices.}
\label{fig:2}
\end{figure}

Both electrons and protons are propagated inside the PWN, and the consequent electron spectra as well as the proton spectra with different spectral indices $\alpha_{\rm p}$ are shown in Fig. \ref{fig:2}. In our calculations, $\alpha_1=1.5$ and $\alpha_2=2.5$ are used for electrons while three different spectral indices are adopted for protons($\alpha_{\rm p}=2.0,~2.1,~2.2$). As illustrated in the plot, the total electron spectrum is dominated by primary electrons, while the contribution by secondary electrons produced by p-p interactions is negligible.

\begin{figure}
\figurenum{3}
\centering
\includegraphics[width=0.55\textwidth,angle=0]{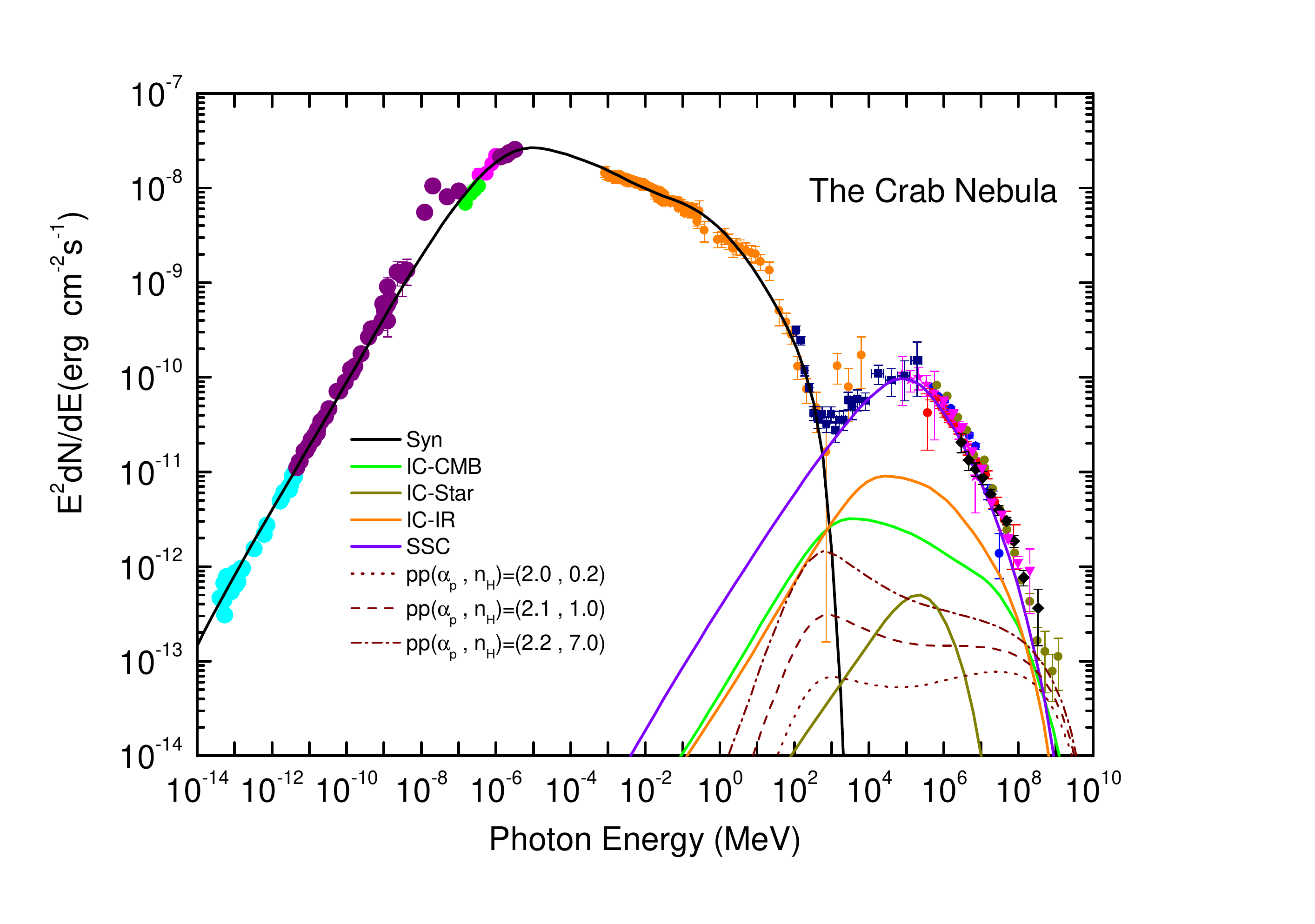}
\caption{The SEDs of the Crab Nebula are plotted (see legends in the figure), where both leptonic and hadronic contributions are considered. The adopted parameters are listed in Table 1. For a direct comparison, observed multi-band data are also displayed. The observed data are taken from \citet{Baldwin1971} and \citet{Macias2010} at radio band, \citet{NS1968}, \citet{Grasdalen1979}, and \citet{Temim2006} at IR band, \citet{Veron-Cetty1993} at optical band, \citet{Hennessy1992} and \citet{Kuiper2001} at X-ray and soft-$\gamma$ ray bands, and
\citet{Aharonian2004,Aharonian2006}, \citet{Albert2008}, \citet{Abdo2010}, \citet{Amenomori2019}, and \citet{LHAASOCrab} at $\gamma$-ray band.}
\label{fig:3}
\end{figure}

As mentioned above, in our lepto-hadronic model, the non-thermal photons from the Crab Nebula consist of two distinct components: the leptonic and hadronic contributions. In Fig. \ref{fig:3}, for both electrons and protons, the separate SEDs arising from different radiation processes as well as the total SED are presented. The synchrotron radiation dominates non-thermal emission at \textbf{low energies} up to several MeVs (labeled as Syn). Meanwhile, IC scattering dominates at \textbf{high energies} up to $\gtrsim 200$ TeV, in which the energetic electrons scatter off synchrotron photons (labeled as SSC). As for the hadronic contribution, three combinations of $(\alpha_{\rm p}, n_{\rm H})$ are considered here \footnote{for each given $\alpha_{\rm p}$, the desired $n_{\rm H}$ is required to best reproduced the $\sim$PeV emission in the Crab Nebula.}. Since $\gamma$-rays could be produced in $\pi^0$ decay due to p-p interactions, both the spectral index $\alpha_{\rm p}$ and the number density $n_{\rm H}$ of medium within the PWN are vital parameters. In fact, a small value of $\alpha_{\rm p}$ would result in a small $n_{\rm H}$, and vice versa (see Fig. \ref{fig:3}).

\begin{figure}
\figurenum{4}
\centering
\includegraphics[width=0.55\textwidth,angle=0]{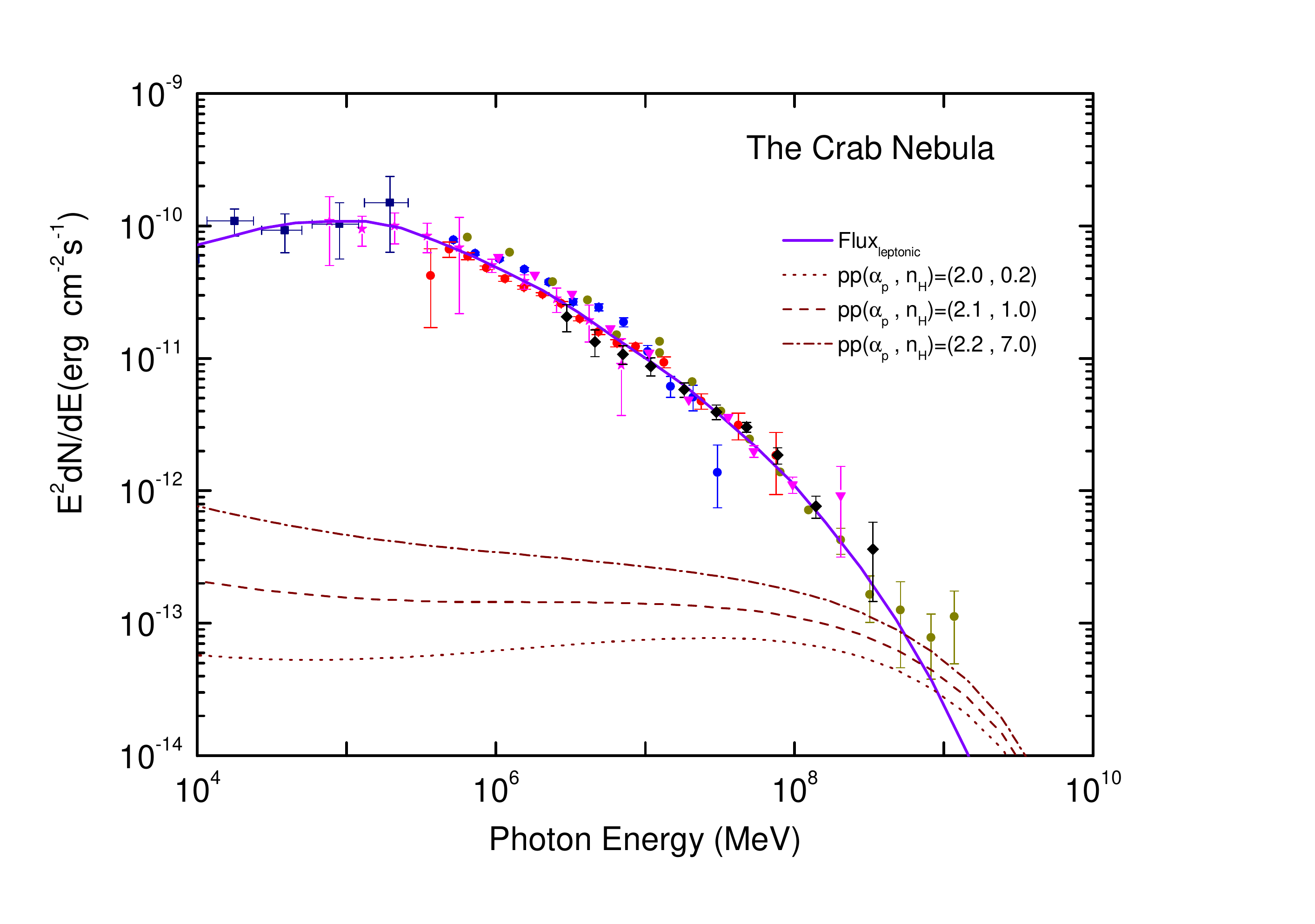}
\caption{\textbf{A zoomed-in plot of Figure \ref{fig:3} where the range of photon energy is from $10^4$ MeV to 10 PeV. The solid line stands for the summed leptonic contributions while the dashed (dotted and dot-dashed) dark red lines are proton spectra with different spectral indices.}}
\label{fig:4}
\end{figure}

To be specific, all three combinations considered here can make significant contributions to $\gamma$-rays with photon energy $\gtrsim 200$ TeV. Fig. \ref{fig:4} illustrates the predicted SEDs between $10^4$ MeV and $\sim$ PeV, where the summed leptonic contributions and individual hadronic components (three combinations of $(\alpha_{\rm p}, n_{\rm H})$ ) are displayed. As shown in the plot, the hadronic components are important for photon energy $\gtrsim 200$ TeV, and our results are consistent with the observed data.

\begin{figure}
\figurenum{5}
\centering
\includegraphics[width=0.55\textwidth,angle=0]{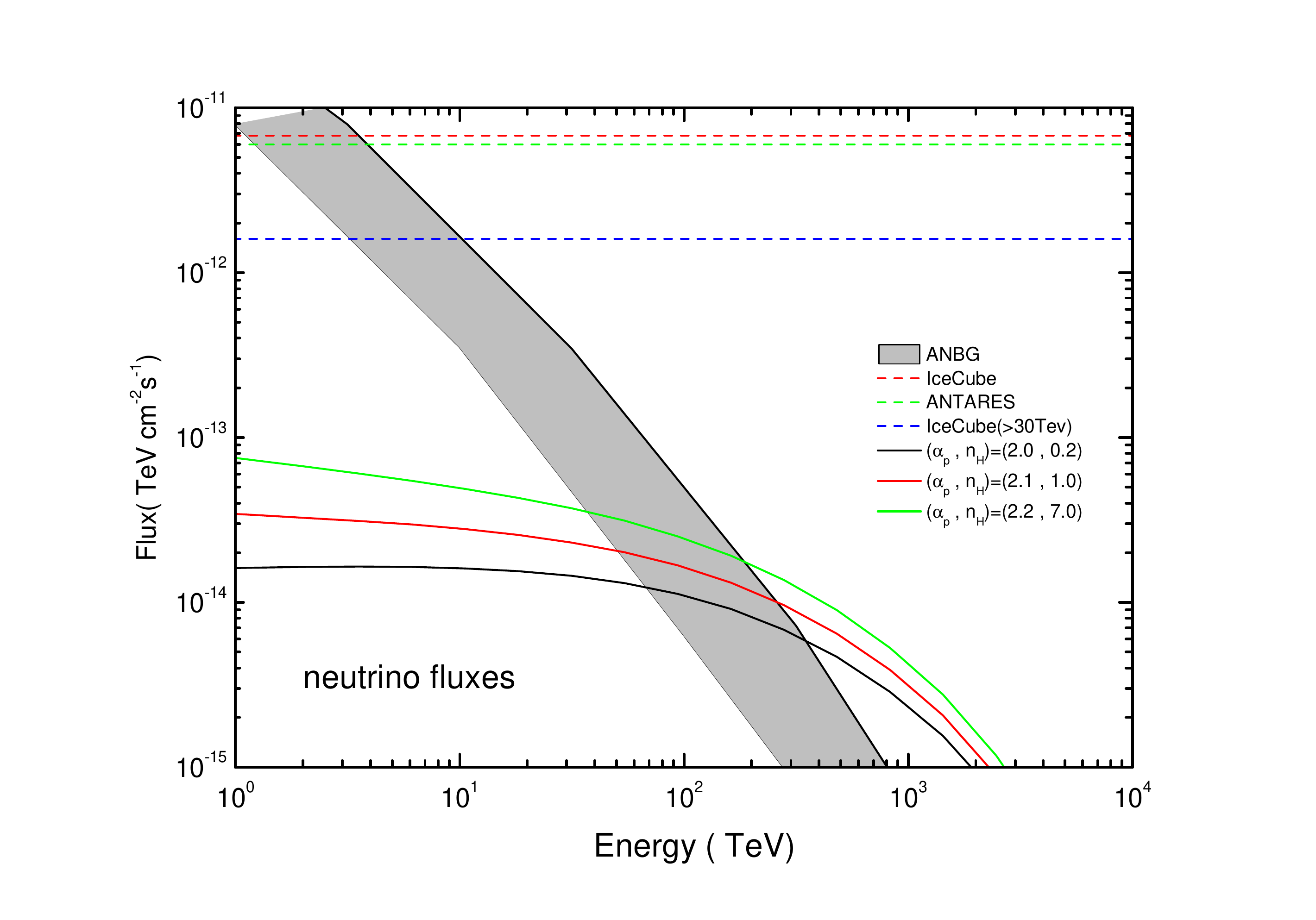}
\caption{Predicted neutrino fluxes from the Crab Nebula for different combinations of parameters $(\alpha_{\rm p}, n_{\rm H})$ in the framework of lepto-hadronic model. For comparison,  the atmospheric neutrino background (ANBG) given by \citet{Adrian-Martinez2013} and
the sensitivities of the IceCube  \citep{Aartsen2017} and ANTARES \citep{Albert2017} are also shown.}
\label{fig:5}
\end{figure}

\begin{figure}
\figurenum{6}
\centering
\includegraphics[width=0.55\textwidth,angle=0]{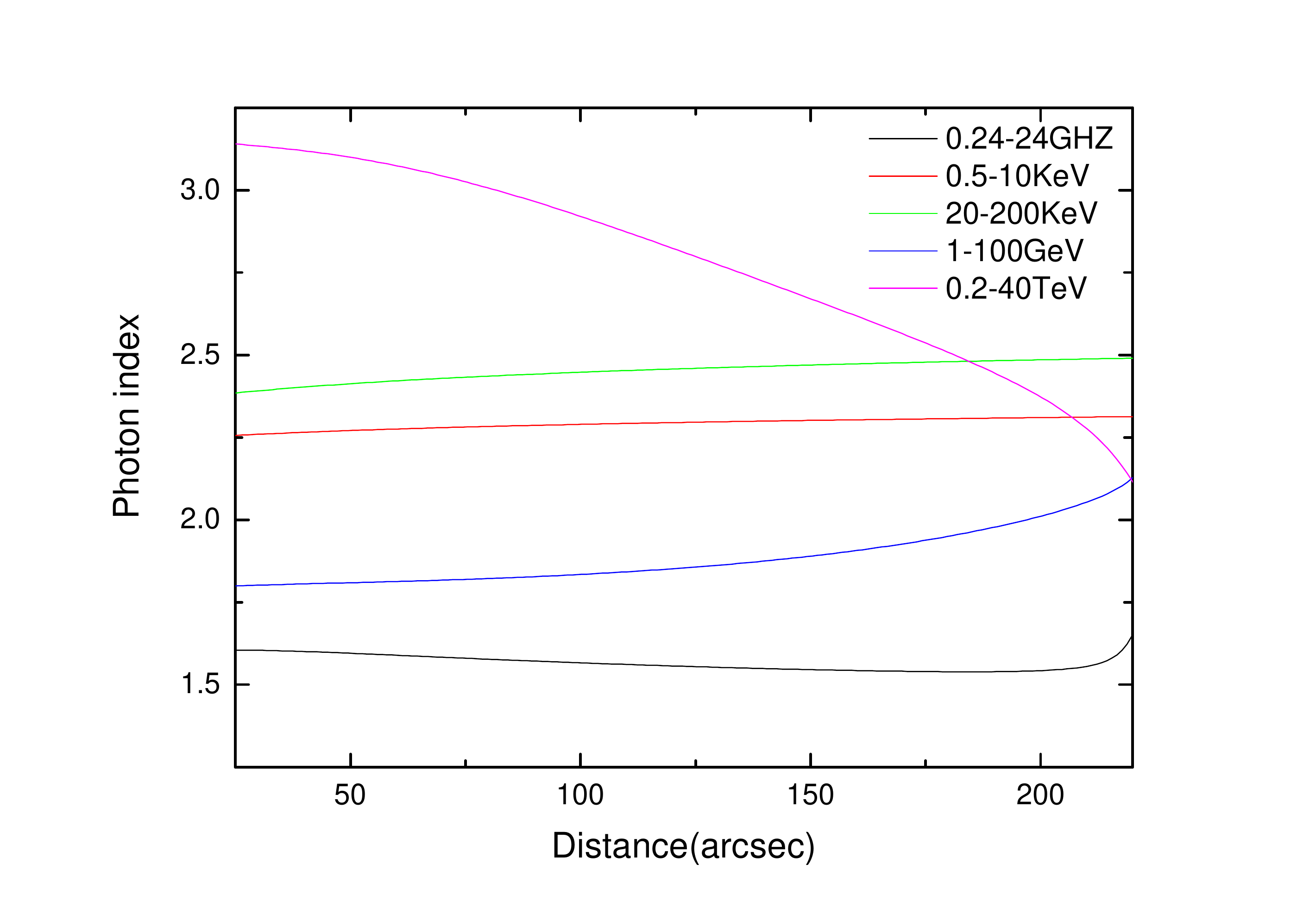}
\includegraphics[width=0.55\textwidth,angle=0]{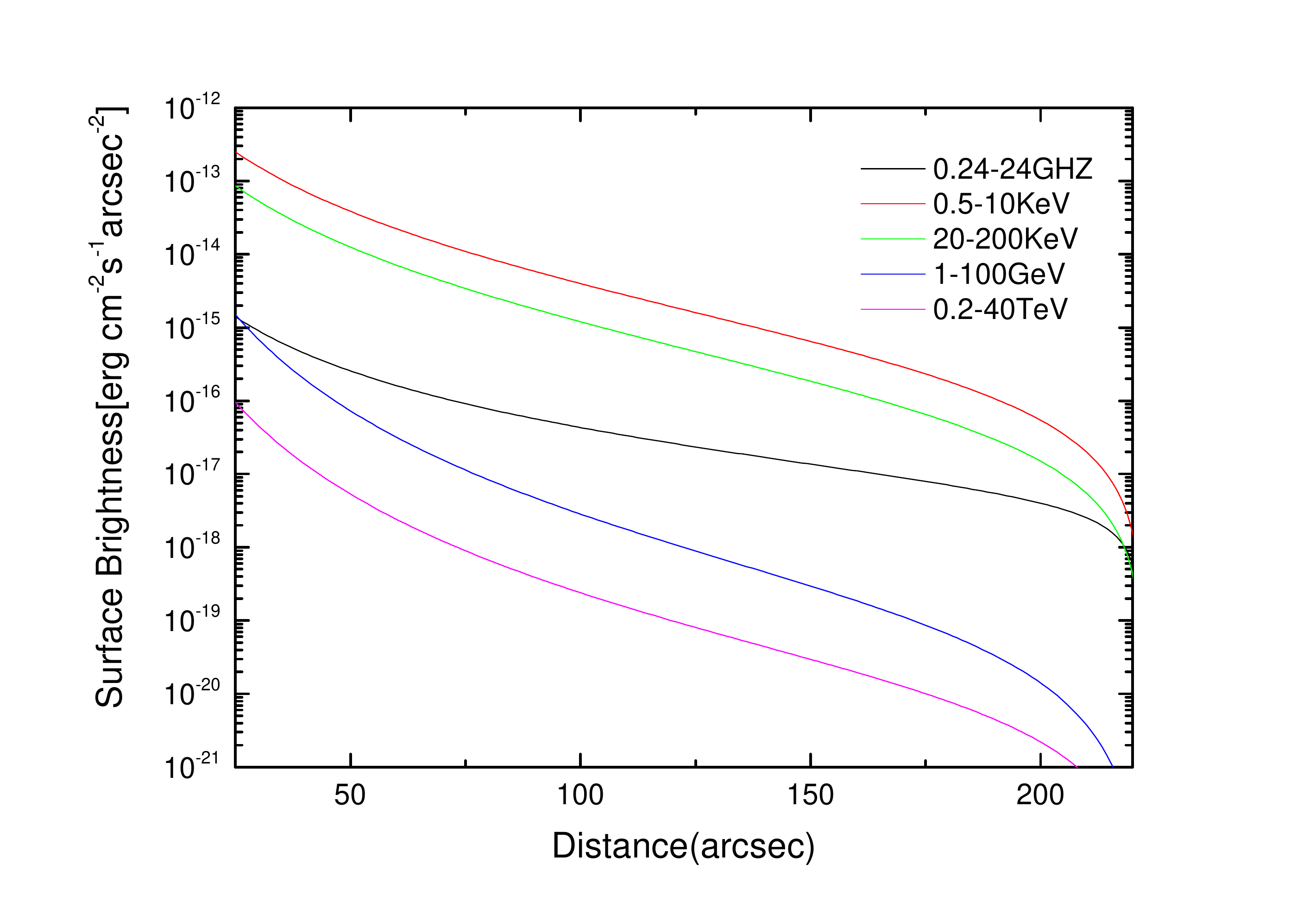}
\caption{Variations of the spectral index (upper panel) and surface brightness (bottom panel) of the Crab Nebula. The different colored lines correspond to the results in different energy ranges.}
\label{fig:6}
\end{figure}

\begin{figure}
\figurenum{7}
\centering
\includegraphics[width=0.55\textwidth,angle=0]{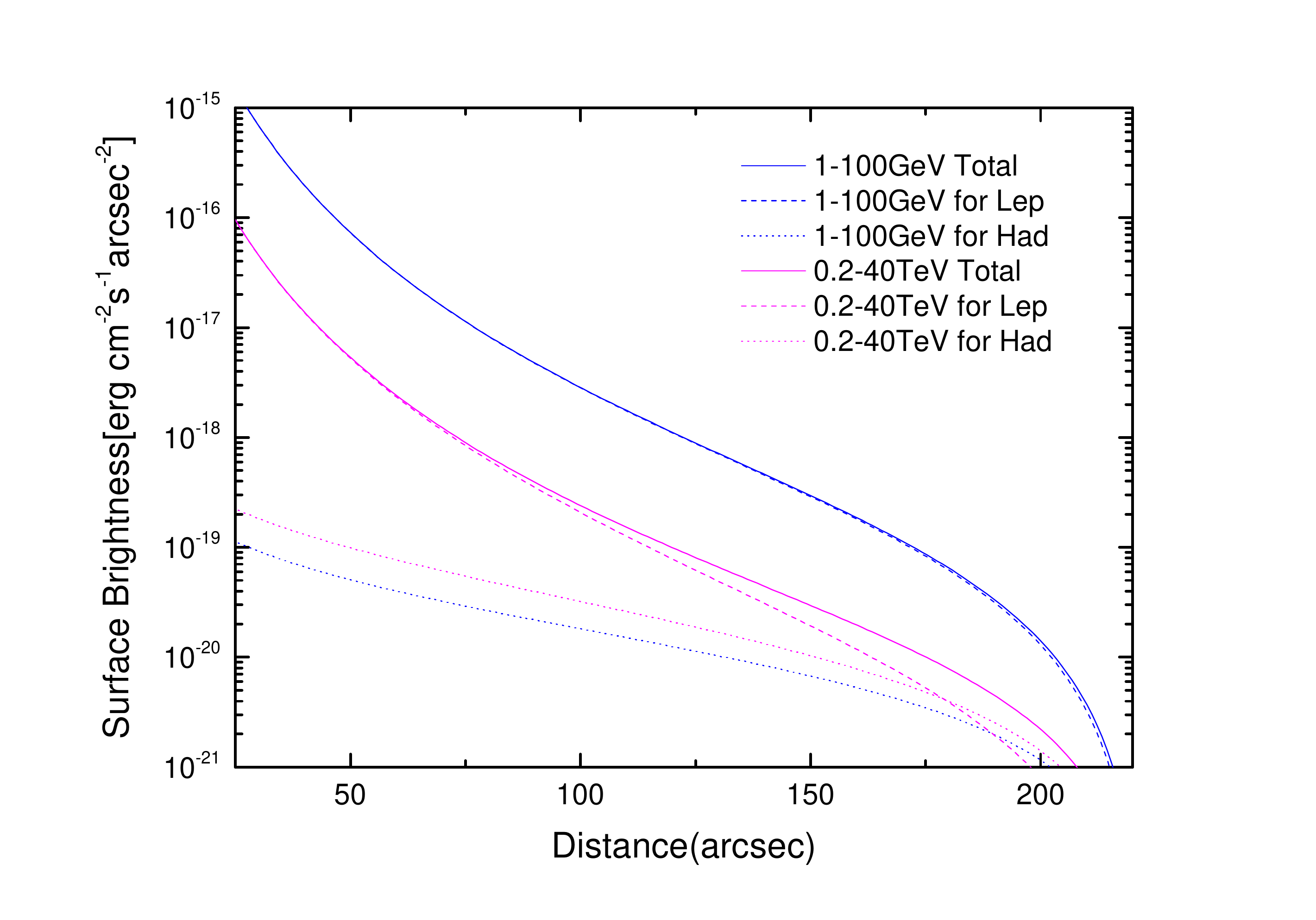}
\caption{Variations of the surface brightness of the Crab Nebula. The blue lines refer to the results at 1-100 GeV while the magenta lines represent the results at 0.2-40 TeV. The solid lines correspond to total surface brightness, while the dashed and dotted lines represent the contributions from leptonic and hadronic components, respectively.}
\label{fig:7}
\end{figure}

In our lepto-hadronic model, TeV neutrinos are produced via hadronic processes. For the Crab Nebula, by using Eq. (\ref{neutrinoplux}), the predicted neutrino fluxes with three combinations of $(\alpha_{\rm p}, n_{\rm H})$ are calculated and plotted in Fig. \ref{fig:5}, where the atmospheric neutrino background (ANBG,within $1^\circ$) given by \citet{Adrian-Martinez2013} is considered. Besides, the sensitivities of the IceCube \citep{Aartsen2017} and the ANTARES \citep{Albert2017} are also shown. It is evident that, our predicted fluxes are not only far below the sensitivities of current neutrino observatories, but also beneath the atmospheric neutrino background with energy less than $\sim 40$ TeV.

Fig. \ref{fig:6} describes the radial profiles of surface brightness and spectral index, where each distinct energy range is marked with a specific colored line.
The upper panel displays the variations of spectral index with the increasing radial distance. As is clearly depicted, in the energy ranges of 0.24-24 GHz, 0.5-10 keV, 20-200 keV and 1-100 GeV, the variations of spectral indices are slight, except in the region close to the outer boundary of the PWN. For the TeV band (0.2-40 TeV), the spectral index manifests an evident decline with the increasing radial distance. The bottom panel shows the surface brightness. As can be seen from the plot, in all energy ranges, the surface brightness decreases with the increasing radial distance. However, the trends of the decrease differ, as the radio band (0.24-24 GHz) retains a mild decrease when compared with other bands.

Our lepto-hadronic model consists of two distinct components: the leptonic and hadronic contributions. In Fig. \ref{fig:7}, for GeV and TeV bands (1-100 GeV, 0.2-40 TeV), the separate surface brightness arising from each component as well as the total surface brightness are presented. As is displayed, at 1-100 GeV, the total surface brightness is clearly dominated by the leptonic contribution, while the hadronic contribution is marginal. For the TeV band (0.2-40 TeV), the leptonic contribution is still in dominance, except in the region close to the outer boundary of the PWN.

\section{Summary and discussion} \label{sec:summary}

As a young filled-center SNR, the Crab Nebula is a potentially powerful cosmic-ray accelerator. The recent LHAASO detection of PeV $\gamma$-rays in the Crab Nebula may serve as the solid identification of a PeVatron. Despite the fact that the multi-band data are well explained within a pure leptonic framework, the potential contribution from the hadronic process can not be ruled out, as any detected $\gamma$-ray photons well beyond 1 PeV would require a non-leptonic origin \citep{LHAASOCrab}. Thus, the lepto-hadronic model as well as its potential contribution to photon spectrum(especially at PeV energies), merit a careful consideration.

Specifically, in this paper, the properties of multi-band non-thermal photon emission from the Crab Nebula are carefully studied in a spatially-dependent lepto-hadronic model. Particularly, a possible hadronic origin of PeV emission from the Crab Nebula is stressed. Generally, the SED from radio to $\sim 200$ TeV is dominated by the leptonic process. While for photon energy $\gtrsim 200$ TeV, the hadronic contribution is important. Although TeV neutrino fluxes are calculated in our model, the current neutrino observations cannot provide any limit for the combination $(\alpha_{\rm p}, n_{\rm H})$, as our predicted fluxes are far below the sensitivities of neutrino observatories.

In our model, the maximum energy is estimated under the assumption that particle Larmor radius is less than the termination shock radius (see Eq. (\ref{Gammax})). Two parameters are concerned here. One is the magnetic fraction $\eta_{\rm B}$, which is limited to a small value: $\eta_{\rm B}=0.06$ (i.e., $\sigma\sim 0.064$), consistent with the result derived from axisymmetric two-dimensional simulations \citep{DVAB06}. In this case, the energy fraction occupied by the electrons (protons) is $\eta_{\rm e}= 0.70$ ($\eta_{\rm p}= 0.24$). Another parameter is the ratio $\varepsilon$ of the particle Larmor radius to termination shock radius, which is set to be $0.7$ here. Despite that the value of $\varepsilon$ adopted here is slightly larger than those previously used \citep[e.g.,][]{DH92,ZCF2008,Torres2014,ZZF18}, we deem our choice here reasonable.

Besides, the \textbf{aforementioned} combination of $(\alpha_{\rm p}, n_{\rm H})$ plays a key role for the hadronic contribution. From our calculations, a steeper proton spectrum would require a higher density within the PWN. As for photon energy $\gtrsim 200$ TeV, all three combinations considered here can make significant contributions to $\gamma$-rays. It is worth mentioning that, in interpreting the Crab Nebula SED with a spatially-independent model, $n_{\rm H}=10$ cm$^{-3}$ and $\alpha_{\rm p}=2.0$ are commonly used \citep[e.g.,] []{LHAASOCrab}. Meanwhile, the relation between the proton fraction $\eta_{\rm p}$ and the medium density $n_{\rm H}$ within the PWN is approximated as $\eta_{\rm p} < 0.07(n_{\rm H}/10~{\rm cm^{-3}})^{-1}$ (see Model D ($\alpha_{\rm p}=2.0$) of  \citet{ZCHC20}), which gives $n_{\rm H}<2.9$ if $\eta_{\rm p}=0.24$. In fact, similar to $n_{\rm H}$, $\eta_{\rm p}$ could also affect the amplitude of SEDs. However, following the method \textbf{commonly used} in leptonic models, the SED from the synchrotron radiation is calculated by adjusting the values of both $\eta_{\rm e}$ and/or $\eta_{\rm B}$ to reproduce the observed data from radio to about 100 MeV band. When $\eta_{\rm e}$ and $\eta_{\rm B}$ are given, the proton fraction is determined by $\eta_{\rm p}=1-\eta_{\rm e} - \eta_{\rm B}$. Since $\eta_p$ is determined in previous calculations, $n_{\rm H}$ become the primary factor affecting the amplitude of SEDs. Therefore, we mainly discuss the effect of $(\alpha_{\rm p}, n_{\rm H})$ pair here.

Moreover, for the Crab Nebula, the calculated radial profiles of surface brightness and spectral index are presented. For all concerned energy bands, the surface brightness decreases with the increasing radial distance, but the trends of the decrease differ. More specifically, for the GeV and TeV bands (1-100 GeV, 0.2-40 TeV), the separate surface brightness arising from the leptonic or hadronic process is displayed in Fig. \ref{fig:7}. At 1-100 GeV, the total surface brightness is clearly dominated by the leptonic contribution. While at 0.2-40 TeV, the leptonic contribution still dominates the surface brightness, except in the region close to the outer boundary of the PWN. Regarding the spectral index, in all energy bands below TeV, the variations of spectral indices are mild, except in the region close to the outer boundary of the PWN. While in the TeV band (0.2-40 TeV), the spectral index profile manifests an evident decline as the radial distance increases.

Finally, the photon emissivity produced in the hadronic process depends heavily on the medium density $n_{\rm H}$ within the PWN (i.e. target gas density), which is also poorly constrained in PWNe. A simple assumption is that the medium within the PWN is equal to the ISM density \citep[e.g.][]{Horns2006,ZY2009}. However, the propagation of relativistic particles inside the filaments within the nebula could be slower than those in the outside, and then the relativistic particles would be partially captured and accumulated in the dense filaments \citep[e.g.][and references therein]{AA1996}. In this case, the effective density of the target gas for the interactions of the relativistic hadrons may be much higher than the density in the surrounding ISM \citep[][]{AA1996}. Thus, in our lepto-hadronic model, the effective density of the target gas in the nebula is treated as an adjustable parameter and the value is estimated to be $n_{\rm H}=0.2,1.0,7.0~{\rm cm^{-3}}$ to reproduce the TeV $\gamma$-ray spectrum with different proton spectra indices $\alpha_{\rm p}$.

For many well-observed PWNe, the pure leptonic one-zone scenario can well reproduce the observed data, without introducing an additional hadronic component. Specifically, there is considerable agreement \textbf{in literature}, that is, synchrotron radiation primarily dominates from radio to X-ray bands while inverse Compton process is responsible for MeV-TeV emission. However, in terms of PeV energies, the hadronic contribution due to p-p interaction may definitely not be marginal, as any detected $\gamma$-ray photons well beyond 1 PeV would require a non-leptonic origin \citep{LHAASOCrab}. Despite the fact that an increased inverse Compton emission seems to properly explain the data with photon energy $\gtrsim 200$ TeV (i.e. reproducing the synchrotron peak and the ICS peak simultaneously), \textbf{a} realistic assessment of the potential hadronic contribution in ultra-high energy band is our primary motivation here. The data accumulated by current $\gamma$-rays observatories could help to discriminate between different scenarios in the near future. As for the Crab Nebula, our results seem to reveal that the leptonic component is dominated by primary electrons while the contribution by secondary electrons due to hadronic process is ignorable (Fig. \ref{fig:2}). Thus, the decay of charged pions to leptons seems unimportant here. For other parameters related with leptonic process listed in Table 1, we \textbf{adopt} their empirical or observational values as default, since our primary focus here is the potential contribution by hadronic process.

In conclusion, the observed multi-band SEDs of the Crab Nebula can be well interpreted in a spatially-dependent lepto-hadronic model, in particular, the SED with energy $\gtrsim 200$ TeV may originate from the contribution of the hadronic component.

\acknowledgments
We thank the anonymous referee for his/her very constructive comments with patience and carefulness. This work is partially supported by National Key R \& D Program of China under grant No. 2018YFA0404204, and the National Natural Science Foundation of China U1738211.

\end{document}